\documentclass[twocolumn,showpacs,preprintnumbers,amsmath,amssymb]{revtex4}
\usepackage{graphicx}
\usepackage{dcolumn}
\usepackage{bm}
\usepackage{xcolor}

\newcommand{\bef}{\begin{figure}}
\newcommand{\eef}{\end{figure}}

\newcommand{\be}{\begin{equation}}
\newcommand{\ee}{\end{equation}}
\newcommand{\bea}{\begin{eqnarray}}
\newcommand{\eea}{\end{eqnarray}}
\begin{document}

\title{Probing the profile of bulk matter in p+Pb collisions via directed flow of heavy quarks}

\author{Md Rihan Haque$^{1}$,  Subhash Singha$^{2}$, and Bedangadas Mohanty$^{3}$}
\affiliation{$^{1}$Warsaw University of Technology, Warsaw, Poland;\\
$^{2}$Institute of Modern Physics Chinese Academy of Sciences, Lanzhou, Peoples Republic of China;\\
$^{3}$National Institute of Science Education and Research, Jatni, India}

\begin{abstract}
The asymmetric initial geometry in p+Pb collisions provide an opportunity to probe the initial matter distribution in these collisions. Using A Multi-Phase Transport (AMPT) model, we have studied the production of light and heavy quark particles in p+Pb collisions at $\sqrt{s_{\rm NN}}$ = 5.02 TeV. The pseudo-rapidity density (dN/d$\eta$) and elliptic flow ($v_{2}$) of light-quark particles from AMPT reasonably agrees with the measurements by ALICE. We predicted the directed flow of light and heavy quarks in p+Pb collisions. The $v_{1}$ of both light and heavy-quark mesons show a non-trivial $p_{T}$ dependence in p- and Pb-going directions. When integrated over the transverse momentum range 0 $ < p_{T} < $5 GeV/c, the magnitude of heavy quark directed flow ($v_{1}$) is found to be 15 times larger than the light quark species in the Pb-going direction while the same in the p-going direction are comparable. The light and heavy quarks $v_{1}$ may offer the possibility to probe the initial matter density as well as the collective dynamics in p+Pb collisions.
 
\end{abstract}
\maketitle

\section{INTRODUCTION}

Relativistic heavy-ion collision (A+A) experiments are performed to understand the formation and evolution of a strongly interacting matter, called Quark Gluon Plasma (QGP)~\cite{qgp0}. Experiments at the  Brookhaven Relativistic Heavy Ion Collider (RHIC) and at CERN Large Hadron Collider (LHC) facilities have established the existence of such strongly interacting matter~\cite{whitepapers}. Collective motion of the particles (flow) emitted from A+A collisions is of special interest because of it's sensitivity to the equation of state of the system. 
Initially it was assumed that the small system collisions, such as p+A, do not create a QGP medium. However, the recent measurement of high multiplicity events in p+A collisions at both LHC and RHIC energies reveal a strikingly similar collective behavior as those observed in A+A collisions at comparable multiplicities~\cite{alice_v2_pA,alice_v2_pA_muon,cms_flow_pA,phenix_nature_small_sys}. It is still under debate what is the origin of collective behavior in small system collisions. Several hydrodynamics and transport model calculations have shown that such collective behavior in p+A collisions may be attributed to final state interactions similar to A+A collisions~\cite{small_sys_hydro_bozek_2012,small_sys_hydro_bozek_2013,small_sys_trans_bzdak_2014,small_sys_trans_soren_2017}.
The directed flow (called $v_{1}$) quantifies the 1$^{st}$ order azimuthal anisotropy of particles of interest in the momentum space. The magnitude of $v_{1}$ is a response of the initial spatial anisotropy, the expansion dynamics and the equation of state of the medium~\cite{eos1,eos2,eos3}. The $v_{1}$ is defined as~\cite{flow_method,v1review},
\begin{equation}
v_{1} = \langle \rm cos (\phi_{s} - \Psi_{RP}) \rangle,  
\label{eqn1}
\end{equation}
where $\phi_{s}$ denotes the particle azimuthal angle, $\langle ...\rangle$ represents the average at a given rapidity and $\Psi_{RP}$ is the reaction plane angle. 


The heavy quark play a crucial role in probing the QGP. The heavy quarks are produced in hard partonic scatterings following a binary collision profile during the early stages of collisions. Hence the distribution of their production profile is expected to be symmetric with respect to the beam axis. The probability of their thermal production is expected to be small. Due to large mass, they decouple in the early stages of the collision. The total number of charm quarks is frozen quite early in the history of collision. This allows to use heavy quarks to obtain information on early time dynamics. It is pointed out in~\cite{sandeep_Dv1_1,sandeep_Dv1_2} that the heavy quark directed flow ($v_{1}$) can be used as a probe for the longitudinal profile of the bulk matter distribution. Recently the STAR experiment at RHIC observed about 25 times larger magnitude of $v_{1}$ slope for the $D^{0}$-mesons compared to the charged kaons~\cite{D0STAR_v1}. These measurements provide a vital information of the geometry of the matter distribution in the longitudinal direction in heavy-ion collisions, i.e the tilting of bulk medium with respect to the colliding beam direction (z-direction). In a hydrodynamic calculation the rapidity dependence of $v_{1}$ is reproduced by a tilted source~\cite{tilt}. In p+Pb collisions there is an inherent asymmetry in matter distribution along the longitudinal direction. Moreover, the transverse size of the matter distribution is expected to be larger on the Pb-going than the p-going direction. It can result in a stronger collective flow in Pb-going direction. The ALICE collaboration measured the elliptic flow of muons in p+Pb collisions at 5.02 TeV. It is observed that the flow of muons in Pb-going direction is about 16\% stronger than the p-going direction~\cite{alice_v2_pA_muon}. Such an enhancement of flow on the Pb-going side were predicted by hydrodynamic and transport model calculations~\cite{bzdak_forward_backward}.


In this paper, we would like to probe the initial distribution of the bulk medium in p+Pb collisions via the directed flow of light and heavy quark particles. We have used a string melting version (v2.26t9b) of A Multi Phase Transport (AMPT) model~\cite{ampt} (which includes parton coalescence). Approximately 5 million minimum-bias p+Pb collision events at $\sqrt{s_{\rm NN}}$ = 5.02~TeV has been used for the results presented here. Experimental sign conventions of p-going side being the forward (positive) rapidity and Pb-going side is the backward (negative) rapidity has been followed. While the rapidity-odd component of $v_{1}$ ($v_{1}^{\rm odd} (y) = -v_{1}^{\rm odd} (-y)$) is related to the reaction plane, the even component $v_{1}$ ($v_{1}^{\rm even}$) is expected to originate from the fluctuations within the initial-state colliding nuclei and not related to the reaction plane~\cite{v1even}. Typically one utilizes multi-particle correlation method~\cite{v1even-method} to evaluate $v_{1}^{\rm even}$. In our calculations, the reaction plane angle is fixed at $\Psi_{RP}$=0 (described later). We consider only the rapidity-odd component of directed flow and $v_{1}$ implies $v_{1}^{\rm odd}$. We studied both the transverse momentum ($p_{\rm T}$) and rapidity ($y$) dependence of $v_{1}$ for both heavy and light quarks.


This paper is organized as follows. In the section \textrm{II}, we discuss briefly the AMPT model. Section \textrm{III} describes the psuedorapidity distribution (dN/d$\eta$), $v_{1}$ and elliptic flow ($v_{2}$) results from 5.02~TeV p+Pb collisions using AMPT. The section \textrm{IV} presents a summary of the results.

\begin{figure}
\begin{center}
\includegraphics[scale=0.38]{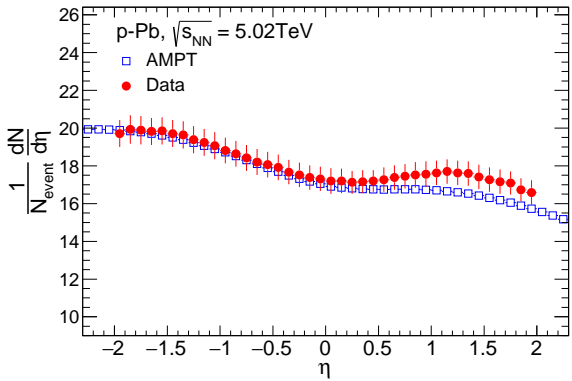}
\caption{Psuedorapidity distribution ($dN/d\eta$ vesrsus $\eta$) of charged particles in p+Pb collisions at $\sqrt{s_{\rm NN}}$ = 5.02~TeV. The blue open squares are results from AMT-SM model and red solid circles are experimental measurements from ALICE collaboration~\cite{alice_dndy_pA}. The vertical lines on the experimental measurements are the uncertainties.}
\label{fig:dndy_ampt_alice}
\end{center}
\end{figure} 

\begin{figure}
\begin{center}
\includegraphics[scale=0.4]{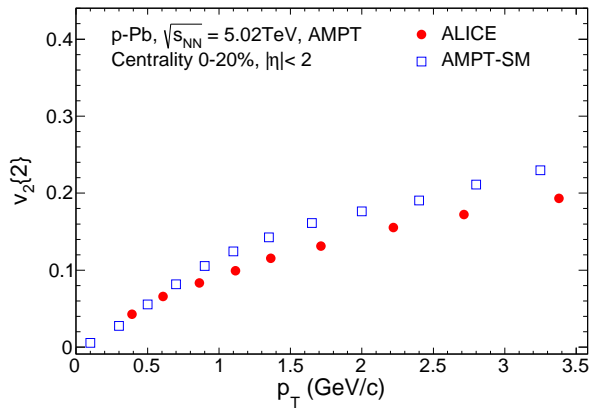}
\caption{Two-particle cumulant method based elliptic flow ($v_{2}$) of charged particles  as a function of transverse momentum ($p_{\rm T}$) in p+Pb collisions at $\sqrt{s{\rm NN}}$ = 5.02 TeV. The open squares are the results from AMPT-SM model and the red solid circles are the corresponding measurements by ALICE Collaboration~\cite{alice_v2_pA}.}
\label{fig-v2pt_ampt_alice}
\end{center}
\end{figure} 

\begin{figure}
\begin{center}
\includegraphics[scale=0.4]{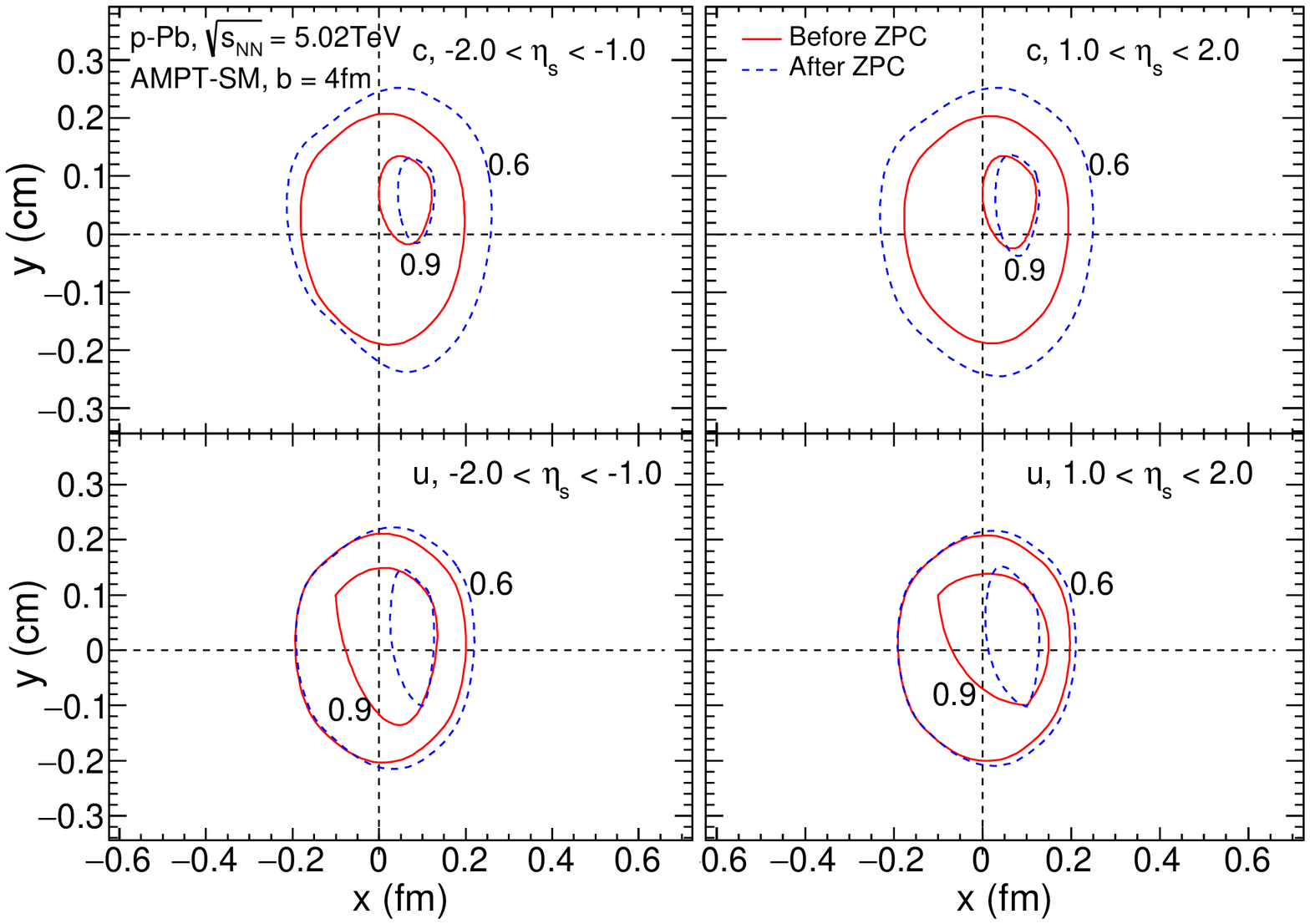}
\includegraphics[scale=0.4]{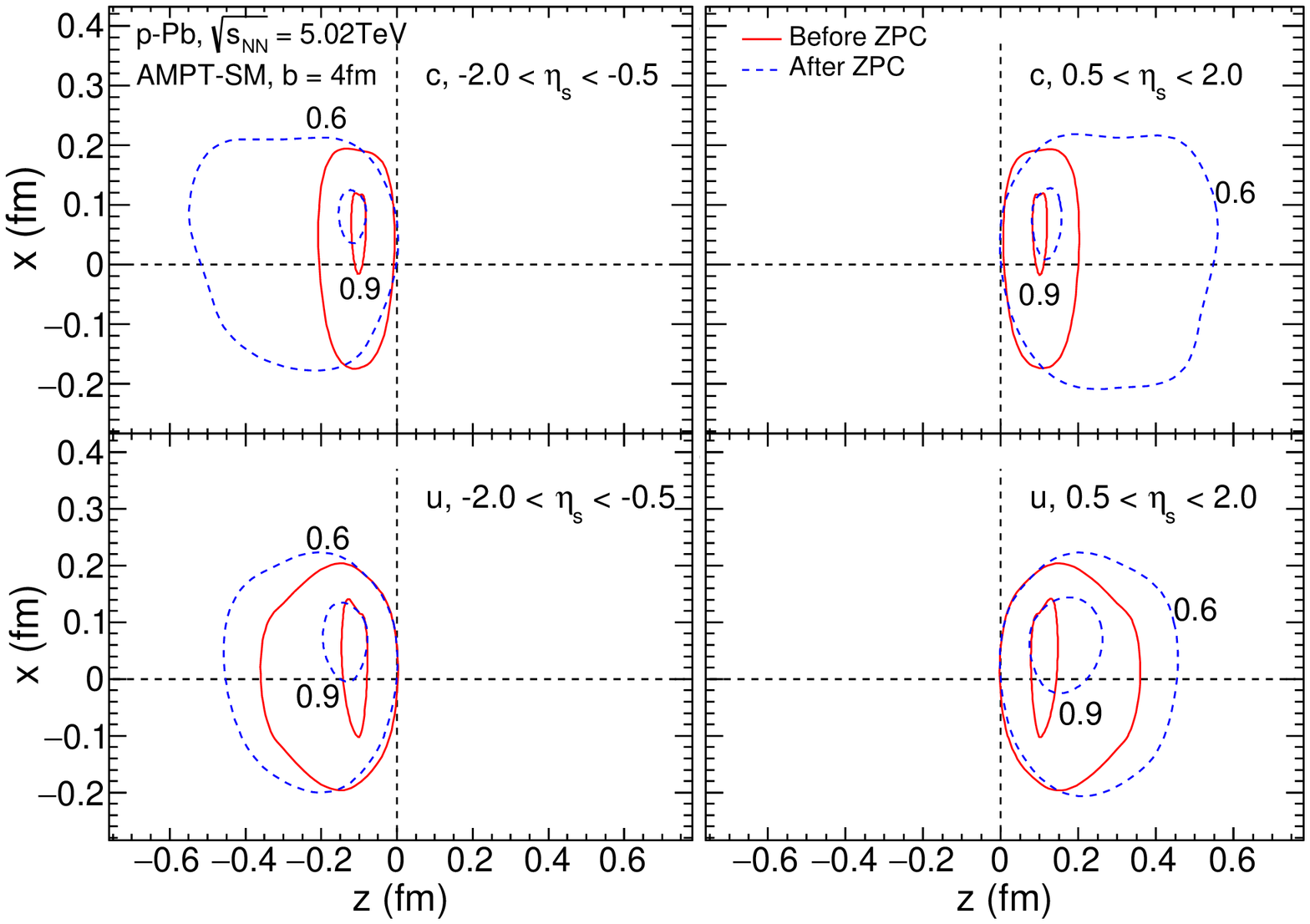}
\caption{The longitudinal and transverse density profile in xz and xy Cartesian frame for light (u) and heavy (c) quarks in p+Pb collisions at 5.02 TeV from AMPT-SM model. $\eta_{s}$ corresponds to spatial pseudorapidity of the quarks. $\eta_{s}>0$ and  $\eta_{s}<0$ corresponds to p-going direction and Pb-going direction, respectively. The $x$, $y$ position of the partons have been recentered event-by-event (see text for details).}

\label{fig:xy-prof-u-c}
\end{center}
\end{figure} 

\begin{figure}
\begin{center}
\includegraphics[scale=0.4]{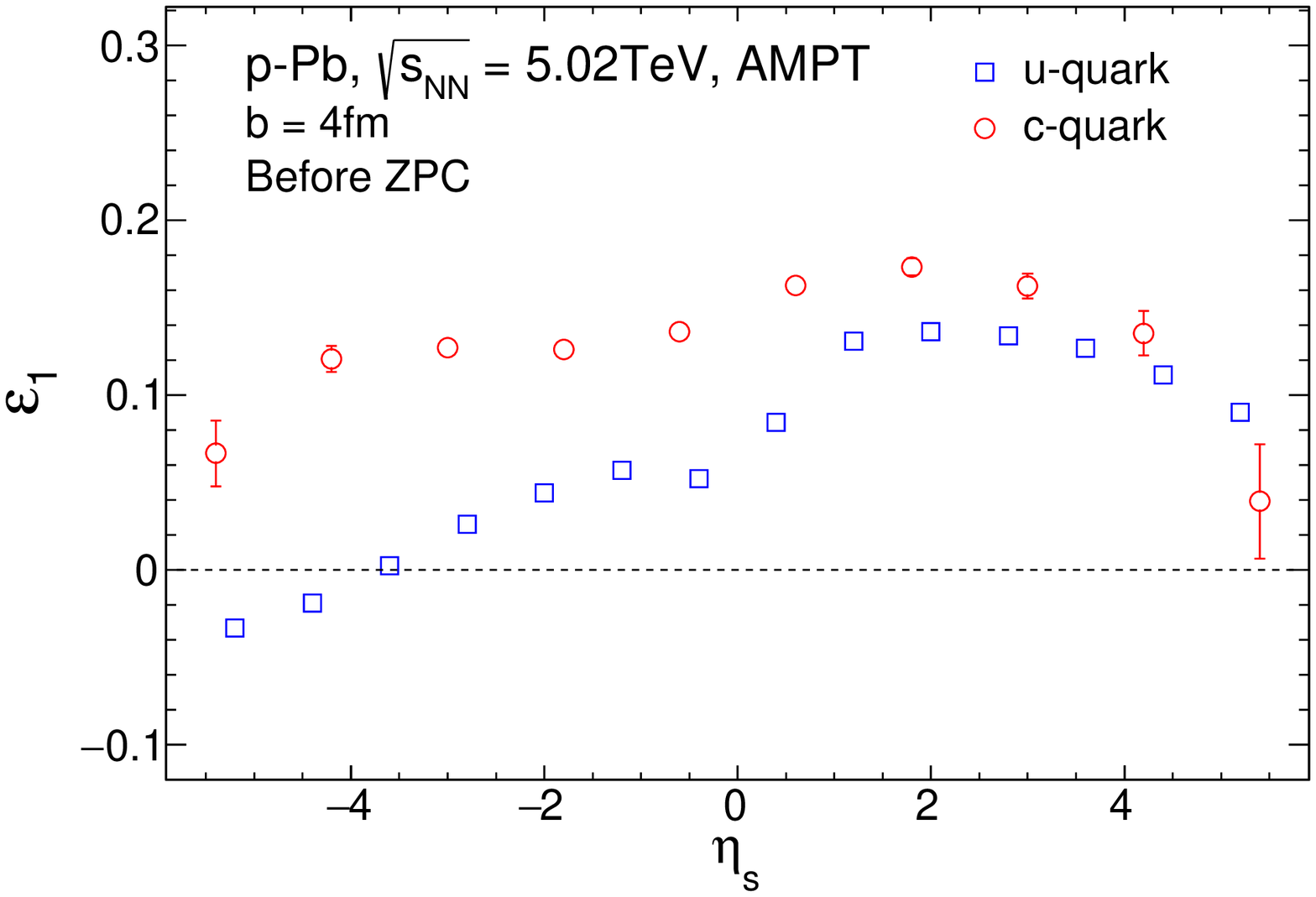}
\includegraphics[scale=0.4]{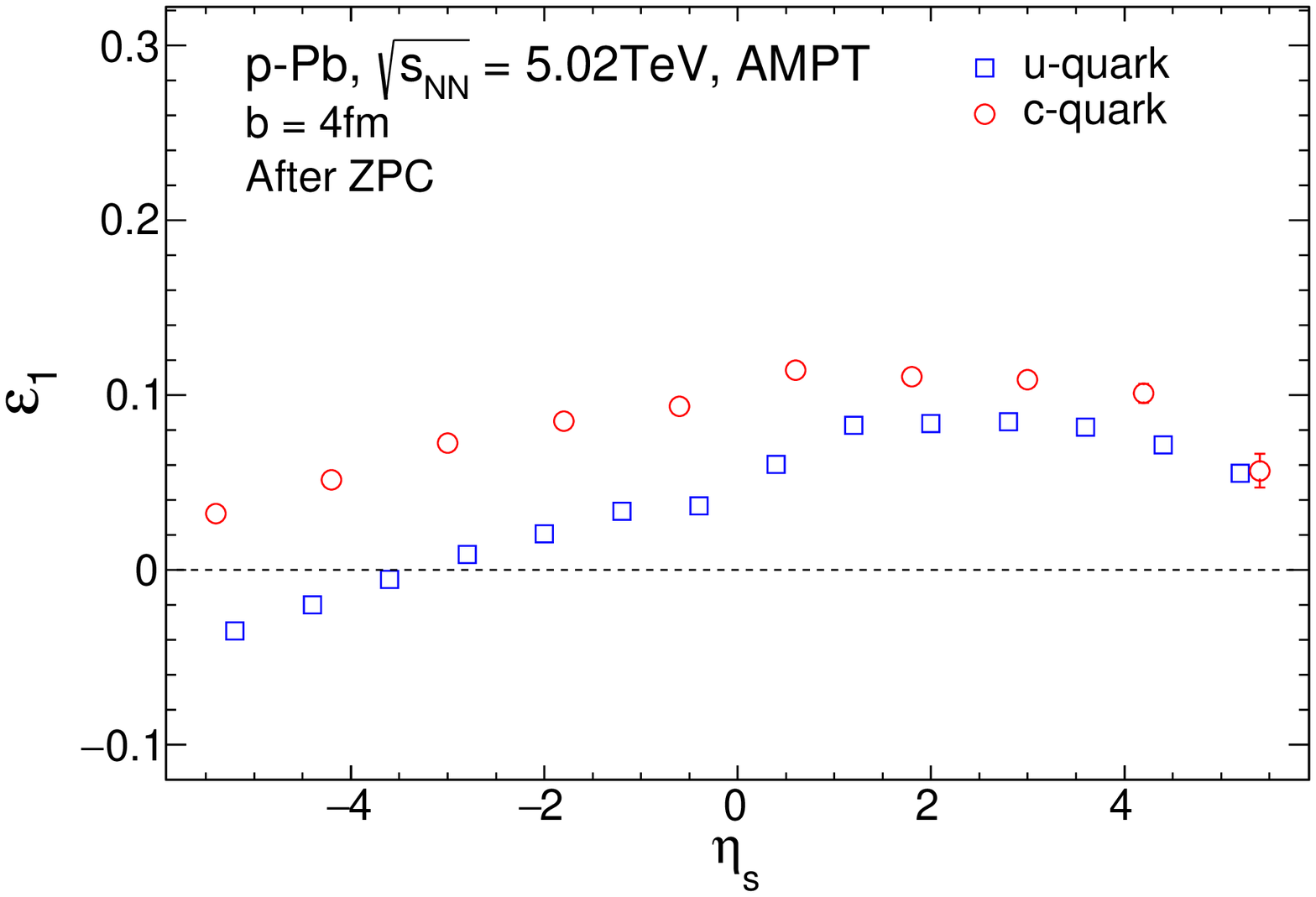}
\caption{Rapidity-odd eccentricity $\epsilon_{1}$ of u and c quarks as function of spatial pseudorapidity $\eta_{s}$ in 5.02 TeV p+Pb collisions for 0 $< p_{T} < $ 5 GeV/c before and after the ZPC in AMPT-SM model.}
\label{fig:epsilon-u-c}
\end{center}
\end{figure} 

\begin{figure}
\begin{center}
\includegraphics[scale=0.4]{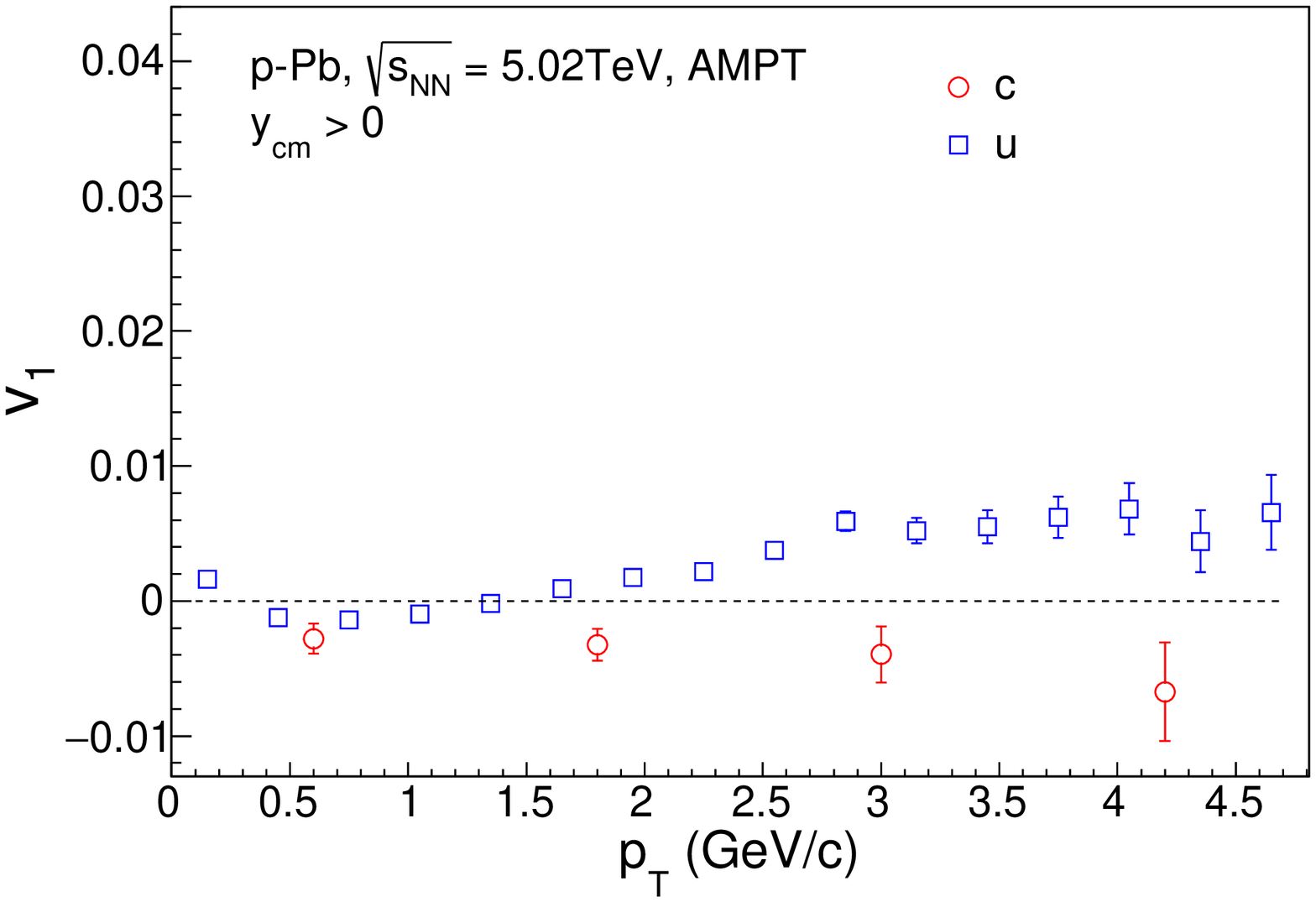}
\includegraphics[scale=0.4]{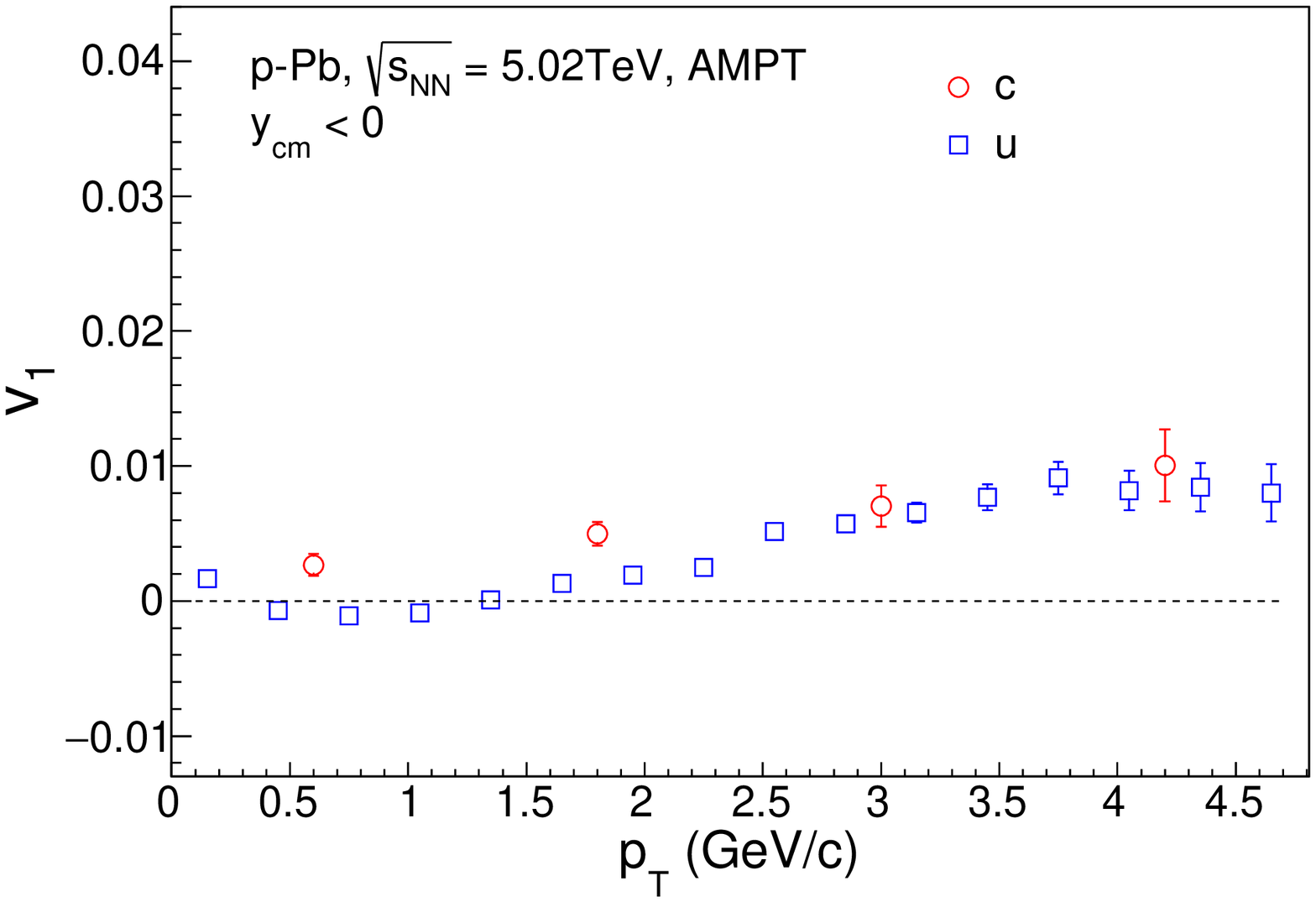}
\caption{Comparison of $v_{1}(p_{\rm T})$ for c and u quarks in forward (top) and backward (bottom) rapidity. Results are from p+Pb collisions at $\sqrt{s_{\rm NN}}$ = 5.02~TeV using AMPT-SM model.} 
\label{fig:parton_v1_pt}
\end{center}
\end{figure} 

\begin{figure}
\begin{center}
\includegraphics[scale=0.4]{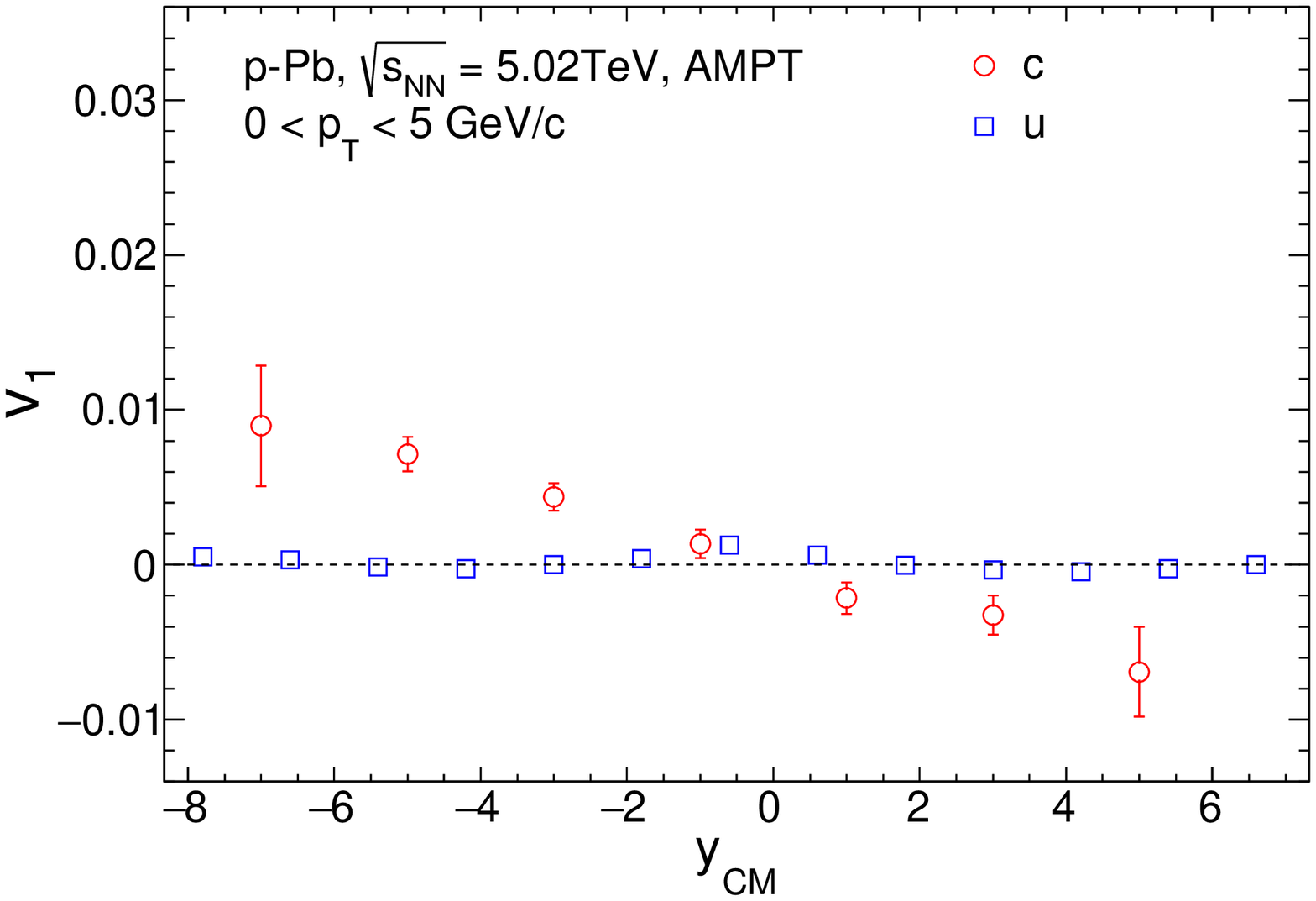}
\caption{Comparison of $v_{1} (y_{\rm CM} )$ for c and u quarks in $p_{\rm T}$ range  0--5~GeV/c. Results are from p+Pb collisions at $\sqrt{s_{\rm NN}}$ = 5.02~TeV using AMPT-SM model.}
\label{fig:parton_v1_rap}
\end{center}
\end{figure} 

\section{The AMPT Model}

The AMPT is a hybrid transport model~\cite{ampt}. It consists of four main components: the initial conditions, partonic interactions, the conversion from the partonic to the hadronic matter, and hadronic interactions. It uses the initial conditions from Heavy Ion Jet Interaction Generator (HIJING)~\cite{hijing}. In the string melting scenario (labeled as AMPT-SM) the excited strings from HIJING are converted to soft partons.
The scattering among partons are modelled by Zhang's parton cascade~\cite{ZPC}, which calculates two-body parton scatterings using cross sections from pQCD with screening masses.  Once partons stop interacting (called parton freezeout later in the text), a quark coalescence model is employed to combine parton into hadrons. The subsequent hadronic interactions are described by a hadronic cascade, which is based on the ART model~\cite{ART}. The parton-parton interaction cross section ($\sigma_{pp}$) in the string-melting version of the AMPT is given by
\begin{equation}
\sigma_{pp} = \frac{9 \pi \alpha_{S}^{2}}{2 \mu^{2}}
\end{equation}
For this study we set the strong coupling constant $\alpha_{S}$ = 0.33 and the parton screening mass $\mu$ = 2.265 fm$^{-1}$. This leads to  $\sigma_{pp}$ = 3 mb. A partonic cross-section of 1.5--3 mb is sufficient to reproduce the p+Pb observations at the LHC energies~\cite{bzdak-ma-parton}. 


\section{Results and Discussion}

The AMPT model with choice of parameter settings as discussed above fairly well reproduce the measured pseudorapidity and elliptic flow in p+Pb collisions at 5.02 TeV. The Fig.~\ref{fig:dndy_ampt_alice} shows the dN/d$\eta$ distribution of charged particles in 5.02~TeV p+Pb collisions from AMPT-SM model. Model results are compared to measurement by ALICE collaboration~\cite{alice_dndy_pA}. We observe that AMPT-SM moderately captures the feature in the data. The pseudo-rapidity distribution in asymmetric collisions (such as p+Pb) is explained by the fact that each participant source preferably showers particles along its direction of motion~\cite{participant-source}. The Fig.~\ref{fig-v2pt_ampt_alice} presents the elliptic flow ($v_{2}\{2\}$) of charged particles from AMPT-SM and the same is compared to the measurements from ALICE~\cite{alice_v2_pA}. The $v_{2}\{2\}$ is obtained from a 2-particle cumulant (2PC) method which gives the RMS value of the quantity measured~\cite{Add-2PC}. So $v_{2}\{2\}$ measurement is in fact $\sqrt{v_{2}^{2}}$. Although we set $\Psi_{RP}$~=~0 in AMPT-SM calculation, the direction of participant plane is ambiguous and hence 2PC method is suitable for $v_{2}$ estimation, specially for p+Pb collisions. We observed that the AMPT model fairly describes the $p_{\rm T}$ dependence of elliptic flow $v_{2}$ of charged particles measured by ALICE~\cite{alice_v2_pA}. The AMPT calculation slightly over predicts the magnitude of $v_{2}$ because 2PC method is susceptible to non-flow, which is dominant process in p+Pb collisions. It is also possible that AMPT may have more contribution from the non-flow compared to data. 
The directed flow refers to the collective motion of particles within the reaction plane consists of the impact parameter (x-axis) and beam direction (z-direction). To understand the initial geometry in p+Pb collisions in AMPT model, we study the density profile of quarks in yx and xz Cartesian frame  before and after the parton cascade (ZPC) phase of AMPT-SM model. Since p$+$Pb is an asymmetric system, we recenter $x,y$ co-ordinates of each partons as $x' = x - \langle x \rangle$ and $y' = y - \langle y \rangle$, where $\langle x \rangle$ and $\langle y \rangle$ corresponds to event-by-event average of $x$ and $y$ positions of the partons. 
For illustration purpose, we show the density profile for u and c quarks in p+Pb collisions at a fixed impact parameter (b) of 4 fm/c. From Fig.~\ref{fig:xy-prof-u-c}, We observe that the density profile of both u and c quarks are asymmetric along the x direction ($i.e.$ along b). There is a relative shift in c-quarks (more asymmetric along x-axis) with respect to that of light u-quarks. 
To quantify this difference in asymmetry for u and c quarks, we calculate the initial rapidity-odd eccentricity ($\epsilon_{1}$) as a function of center of mass pseudorapidity ($\eta_{s}$) using the equation~\cite{epsilon1_1,epsilon1_2},
\begin{equation}
\epsilon_{1} = \langle \rm cos (\phi_{s} - \Psi_{RP}) \rangle,  
\label{eqn3}
\end{equation}
The Fig~\ref{fig:epsilon-u-c} presents $\epsilon_{1}$ as function of spatial pseudorapidity ($\eta_{s}$) for u and c quarks, before and after the ZPC, for p+Pb collisions at impact parameter b = 4 fm. We observe a non-zero magnitude of $\epsilon_1$ for both u and c quarks before the ZPC in both p and Pb-going directions. After the end of partonic interactions, the magnitude of $\epsilon_{1}$ is reduced for both the u and c quarks. However, we observe that the $\epsilon_{1}$ is order of magnitude ($\sim$3.92 times at $\eta_{s}\approx-2.0$) higher for c quark compared to u quark in the Pb-going direction. In the p-going side, the $\epsilon_{1}$ of c quark is slightly larger (32\% at $\eta_{s}\approx2.0$) than the u-quark. Since the $v_{1}$ is a response to initial spatial anisotropy, the observed difference in $\epsilon_{1}$ can cause the difference in $v_{1}$ for u and c quarks.
We also observed that the slope of the $p_{T}$ spectra is harder for c quarks compared to u quarks and the $p_{T}$ slope does not change appreciably before and after the ZPC. 

Next we calculate the $v_{1}$ in minimum-bias p+Pb collisions at $\sqrt{s_{\rm NN}}$=5.02 TeV using the AMPT-SM. For $v_{1}$ we measure the shift of the medium, and hence we use a particle average as described in Eq.~\ref{eqn1}. Recently, in a completely different approach using a hydrodynamic model~\cite{sandeep_Dv1_1,sandeep_Dv1_2} for a symmetric Au+Au collisions, it is pointed out that the shift between the bulk medium and binary collision profile of heavy quarks can induce a huge directed flow for the heavy quarks. Similar feature were also observed in AMPT-SM calculations~\cite{nsm-sbh} done for symmetric Au+Au collisions at RHIC energies. The measurement of large magnitude of $D^{0}$ directed flow compared to the light mesons in Au+Au collisions by the STAR collaboration~\cite{D0STAR_v1} confirms such a scenario. 

In order to address these aspects in p+Pb collisions we study the $v_{1}$ of u and c quarks in p+Pb collisions. The Fig.~\ref{fig:parton_v1_pt} presents $p_{\rm T}$ differential $v_{1}$ for u and c quark in the forward ($y>0$, p going) and backward rapidity ($y<0$, Pb-going) region. We observe a non trivial $p_{\rm T}$ dependence of $v_{1}$ for both u and c quarks in forward and backward rapidity regions. The qualitative feature of $p_{\rm T}$ dependence of $v_{1}$ for u quarks are similar at both forward and backward rapidity regions. While the c quark's $v_{1}$ has a different trend at forward and backward rapidities. This is an interesting observation. We also observe that the 
$p_{\rm T}$ spectra of pions (and u quark) are much softer compared to that of $D^{0}$ meson (and c quark).
The $\langle p_{\rm T} \rangle$ for $D^{0}$ is 2.19 GeV/c while the same for pions is 0.636 GeV/c. 
The observed non trivial $p_{T}$ dependence of $v_{1}$ for u and c quarks may indicate a convoluted effect of initial geometry and different slope of $p_{\rm T}$ spectra. 

The Fig.~\ref{fig:parton_v1_rap} presents the $v_{1}$ as a function of center of mass rapidity ($y_{\rm CM}$) for 0 $< p_{T} < $ 5 GeV/c for u and c quarks after ZPC ($i.e.$ after partonic interactions). We observe that the $v_{1}$ of c quarks is orders of magnitude larger than that of the u quarks in the Pb-going direction. Since the $\langle p_{\rm T} \rangle$ of c quarks is large than u quarks, the rapidity dependence of $v_{1}$ of u quarks predominantly comes from the low $p_{\rm T}$ part of the spectra. The different $v_{1}(y)$ trend of u and c quarks may possibly due to the shift in their transverse density profile (shown in Fig.~\ref{fig:xy-prof-u-c}) combined with their $p_{\rm T}$ spectra. 


\begin{figure}
\begin{center}
\includegraphics[scale=0.4]{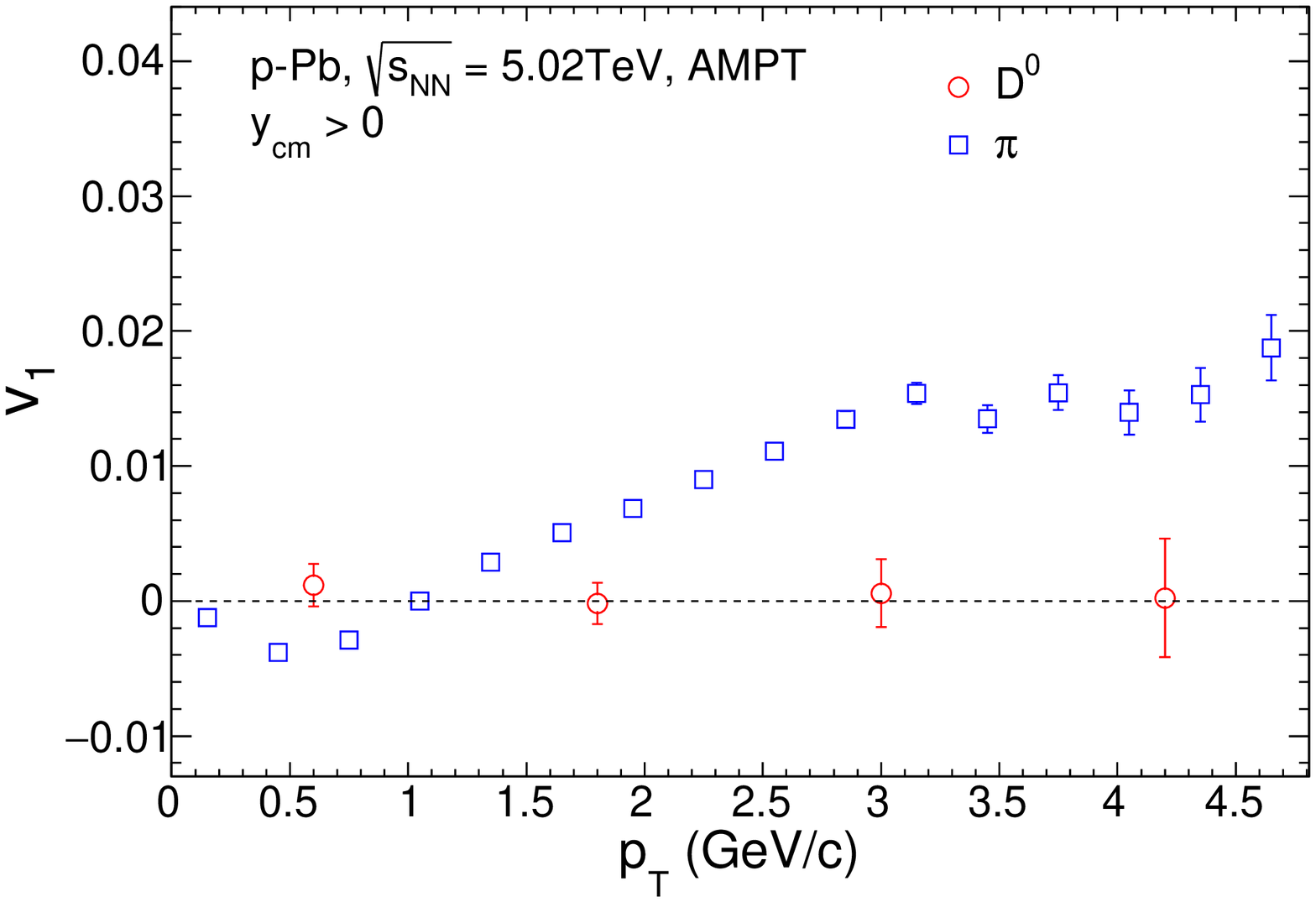}
\includegraphics[scale=0.4]{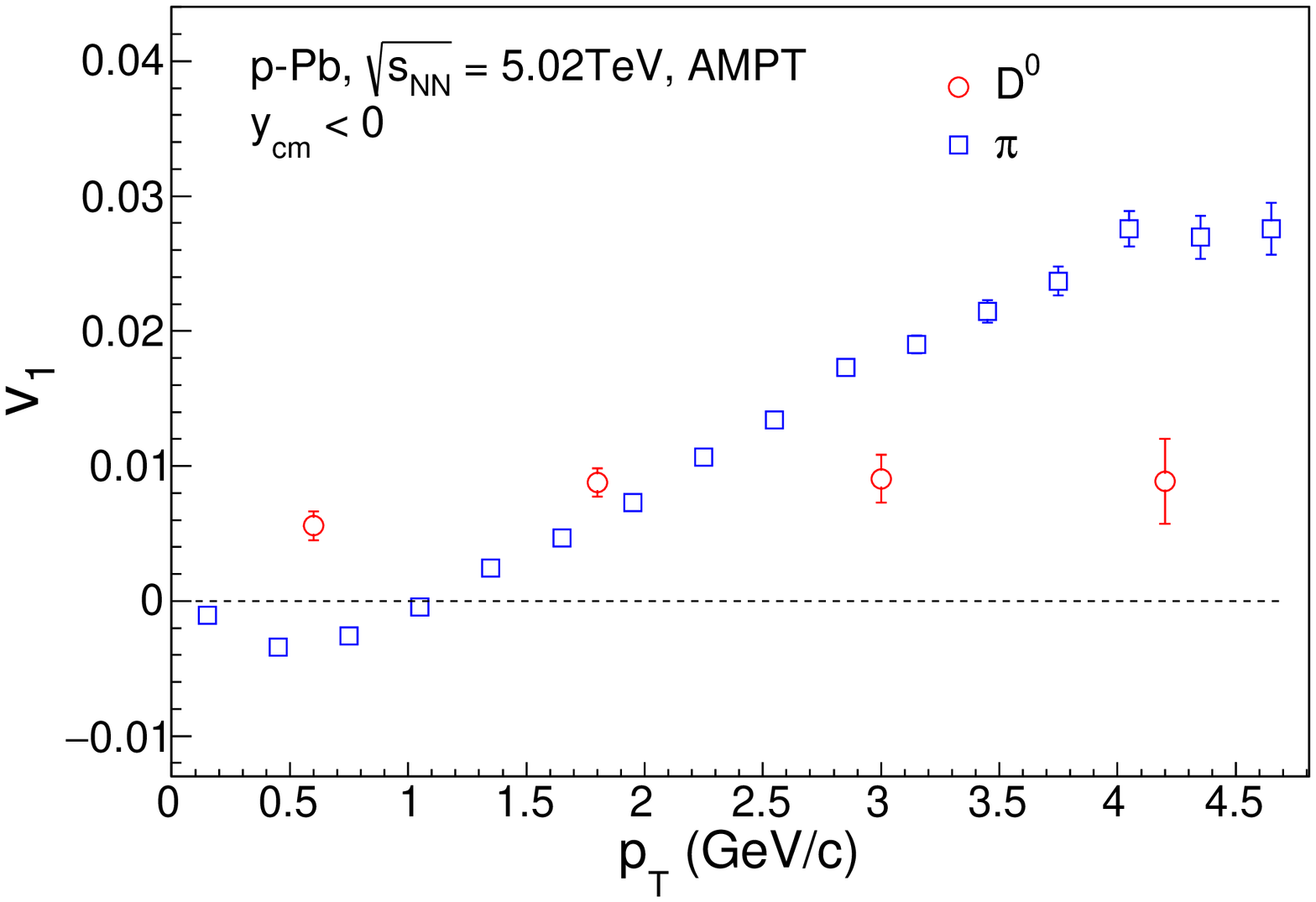}
\caption{Comparison of $v_{1} (p_{\rm T})$ in forward rapidity (top) and backward rapidity (bottom) for $D^{0}$ and $\pi$ in 5.02~TeV p+Pb collisions using AMPT-SM model.}
\label{fig:D0_v1_pt}
\end{center}
\end{figure} 

As described in section~\textrm{II}, to form mesons from the quarks after the partonic interactions, a dynamic coalescence mechanism is used in AMPT-SM~\cite{yu-ziwei}. In this mechanism the probability of formation of a meson is calculated based on the overlap of the position-space distributions of the quark anti-quark pair with the phase space distribution of the meson (in its rest frame)~\cite{yu-ziwei}. Following this method, the u and $\bar{d}$ quarks coalesce to form pions, while c and $\bar{u}$ quarks coalesce to form the $D^{0}$ mesons. After the hadronization, these mesons participate in subsequent late stage hadronic interactions described by the ART model~\cite{ART}. The lifetime of hadronic phase in our study corresponds to 30 fm/c which is a default value in AMPT-SM.

\begin{figure}
\begin{center}
\includegraphics[scale=0.4]{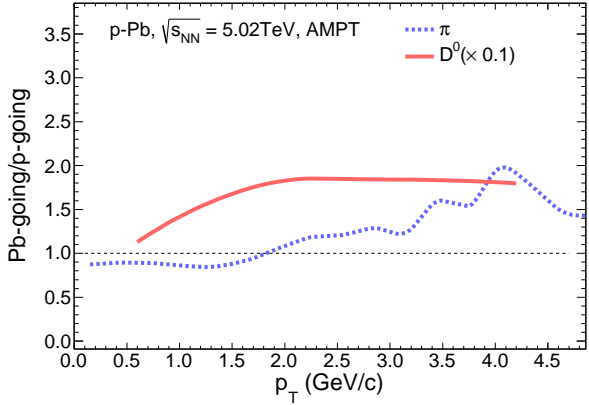}
\caption{Ratio of $v_{1}$ in Pb-going to p-going direction as function of $p_{\rm T}$ for $D^{0}$ (solid line) and $\pi$ (dotted line) in 5.02~TeV p+Pb collisions using AMPT-SM model.}
\label{fig:Ratio_v1_pt}
\end{center}
\end{figure}

The Fig.~\ref{fig:D0_v1_pt} presents $p_{\rm T}$ differential $v_{1}$ for $D^{0}$ and $\pi$'s in the forward ($y>0$) and backward ($y<0$) rapidity region. At both forward and backward rapidity, the pions have a strong $(p_{\rm T})$ dependent $v_{1}$. For $D^{0}$ mesons, the $v_{1}$ is consistent with zero in the forward rapidity and comparable with $\pi$ at the backward rapidity. This feature is qualitatively similar to the behavior of constituent quarks which is shown in~Figure~\ref{fig:parton_v1_pt}. We observe that the magnitude of pion and $D^{0}$ $v_{1}$ is larger than that of u and c quark $v_{1}$. It possibly comes from the hadronization by coalescence and a subsequent late stage interaction in AMPT. We also observe that the magnitude of $v_{1}$ is different in Pb- and p-going direction. To quantify the difference, we present the ratio of $v_{1}$ in Pb-going to the p-going direction as function of $p_{\rm T}$ in Fig~\ref{fig:Ratio_v1_pt}. Since $v_{1}$ of $D^{0}$ in the p-going direction is $\sim$~0, we fit the points with a constant. The fit results into $v_{1}(y>0)$ = 0.00049 $\pm$ 0.00098 for the $D^{0}$. We then make the ratio of $v_{1}$ for $D^{0}$ in the Pb-going direction with this fit value. For pions, the ratio has a non-trivial $p_{\rm T}$ dependence. While for $D^{0}$ mesons ratio is 15 times larger than for pions for $p_{T} > $ 1.5 GeV/c. We also observe an enhancement of $D^{0}$ $v_{2}$ (about 20\%, not shown here) in AMPT-SM in the Pb-going side which is of similar order to that observed for muons by ALICE~\cite{D0STAR_v1} experiment.

\begin{figure}
\begin{center}
\includegraphics[scale=0.4]{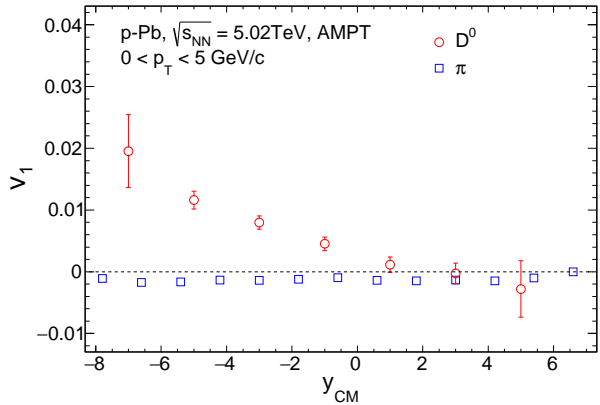}
\caption{Comparison of $v_{1}(y_{\rm CM})$ for for $D^{0}$ and $\pi$ in the $p_{\rm T}$ interval 0--5~GeV/c in 5.02~TeV p+Pb collisions using AMPT-SM model.}
\label{fig:D0_v1_rap}
\end{center}
\end{figure} 

The Fig.~\ref{fig:D0_v1_rap} shows the rapidity dependence of $v_{1}$ for $D^{0}$ and $\pi$ for 0 $< p_{T} < $ 5 GeV/c. The values of $D^{0}$ $v_{1}$ is about two orders of magnitude larger than pions in the backward rapidity (Pb-going direction). In the forward rapidity (p-going direction), the $v_{1}$ for both $D^{0}$ and $\pi$ is consistent with zero. The enhanced magnitude of $D^{0}$ meson heavy $v_{1}$ may possibly comes from the shifted profile of the light and heavy quarks (as shown in Fig.~\ref{fig:xy-prof-u-c}) in conjunction with different $p_{\rm T}$ spectra, coalescence mechanism and late stage hadronic interactions. This is an interesting observation suggest that the simultaneous measurement of pions and $D^{0}$ $v_{1}$ can be a useful probe to understand the initial bulk matter distribution in small system (p+Pb) collisions. The ALICE collaboration observed that the $v_{2}$ of muons in the Pb-going direction is about 16\% larger than p-going direction can be qualitatively captured by the AMPT model calculation~\cite{alice_v2_pA_muon,bzdak_forward_backward}. We have shown that the directed flow heavy quark particles in the Pb-going direction is about 15 times larger than the same in p-going direction. Therefore, the measurement of heavy-quark particle's $v_{1}$ have better sensitivity to the early stage medium than the light quark particles. It has been studied for Au+Au collisions in Ref.~\cite{epsilon1_1} that the hadronization mechanisms and late stage hadronic interactions can modify the $v_{1}$. In future it will be interesting to study the sensitivity of $v_{1}$ with different coalescence hadronization mechanisms (such as, coordinate space versus momentum space coalescence) and with different lifetime of hadronic interactions in p+Pb collisions. 

\section{Summary and Conclusion}
In summary, we have presented the directed flow ($v_{1}$) of light and heavy hadrons, and their constituent quarks in p+Pb collisions at $\sqrt{s_{NN}}$=5.02~TeV using a string melting version of the AMPT model. We observe a non-trivial $p_{\rm T}$ dependence of $v_{1}$ for light and heavy quarks. When integrated over 0 $< p_{\rm T} < $ 5 GeV/c, the magnitude of $D^{0}$ meson $v_{1}$ in the Pb-going direction is orders of magnitude larger than that of the pions. The ordering of magnitude of $v_{1}$ for pions and $D^{0}$ mesons in Pb and p going direction is consistent with that of the $\epsilon_{1}$ for light and heavy quarks in the initial stage. The difference in $v_{1}$ for light and heavy hadrons possibly reflects the effect of the shifted transverse profile in the initial stages weighted with their $p_{\rm T}$ spectra. This unique feature of $v_{1}$ in forward and backward rapidities for light and heavy quarks may be used to probe  the distribution of the bulk matter along the longitudinal direction. The $v_{1}$ results may also help understand the nature of collective dynamics in small system collisions. A recent paper~\cite{phsd} predicted that the initial magnetic field can induce a charge dependent $v_{1}$ in p+Pb collisions. The effect of charge dependent splitting for heavy quarks is expected to be larger than for light quarks~\cite{santosh}. In future, one can study these effects using the same AMPT model as discussed in this work.

\noindent{\bf Acknowledgments}\\
Authors would like to thank Sandeep Chatterjee for discussions and providing fruitful suggestions. SS is supported by the Strategic Priority Research Program of Chinese Academy of Sciences (Grant XDB34000000). MRH is supported by the European Union’s Horizon 2020 grant (agreement No 824093). BM acknowledges support from J C Bose Fellowship of DST, Govt. of India. SS and MRH would like to acknowledge hospitality at NISER-Jatni campus where a part of this work has been done.

\end{document}